\documentclass[11pt]{article}
\usepackage{graphicx}
\usepackage{amssymb}

\title{Sliding friction between an elastomer network and a grafted polymer layer: the role of cooperative effects}

\author{T. Vilmin and E. Rapha\"el\\ \\
Laboratoire de Physique de la Mati\`ere Condens\'ee\\
UMR 7125 CNRS and FR 2438 "Mati\`ere et Syst\`emes Complexes"\\
Coll\`ege de France\\
 11, Place Marcelin Berthelot, 75231 Paris Cedex 05, France\\ \\}

\begin{document}

\maketitle

\begin{abstract}

We study the friction between a flat solid surface where polymer chains have been end-grafted and a cross-linked elastomer at low sliding velocity. The contribution of isolated grafted chains' penetration in the sliding elastomer has been early identified as a weakly velocity dependent pull-out force. Recent experiments have shown that the interactions between the grafted chains at high grafting density modify the friction force by grafted chain. We develop here a simple model that takes into account those interactions and gives a limit grafting density $\sigma_{l}$ beyond which the friction no longer increases with the grafting density, in good agreement with the experimental data.

\end{abstract}

\section{Introduction}

\begin{figure}
   \centering
   \includegraphics[clip=true, bb=6.5cm 6.5cm 22cm 13.4cm, width = 8.5cm]{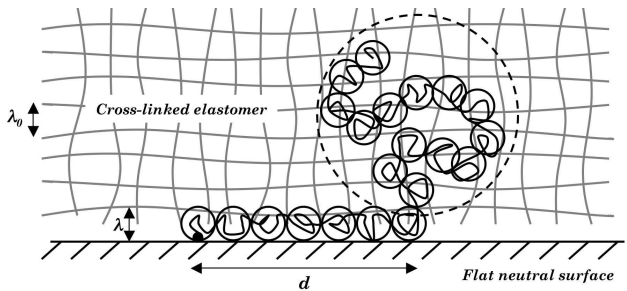}
\caption{Grafted chain that has partially entered the elastomer ($\lambda_{0}$ is the distance between cross-links, $\lambda$ the slab width, and $d$ is the stretching length of the $n_{eq}$ confined monomers). }
\label{f.1}
\end{figure}

In recent years, polymer chains grafted to a surface or an interface
have been the subject of many theoretical and experimental studies
because of their practical importance\cite{Jones}. In particular, the interface between
a solid surface and a crosslinked elastomer network can be strengthened by the
addition of chains that are tethered by one end to the solid surface.
As a crack grows along the interface, these coupling chains are progressively
pulled-out from the elastomer leading to significant energy dissipation\cite{Brown94}\cite{LRH99}.
The presence of end-tethered chains plays also an important role in friction\cite{Bubin93, Brown94,
Ajdari94, LRH99}.
Very recently, Bureau and L\'eger \cite{Bureau04} studied the friction of a poly(dimethylsiloxane) (PDMS)
elastomer network sliding, at low velocity, on a substrate on which PDMS chains are
end-tethered and clearly evidenced the contribution to friction of the pull-out
mechanism of chain-ends that penetrate into the network. This study, while confirming semi-quantitatively the picture of arm retraction relaxation
of the grafted chains proposed by Rubinstein \textit{et al.} \cite{Bubin93}\cite{Ajdari94}, also reveals the unexpected feature that the friction stress, after increasing with the grafting
density of tethered chains, reaches a plateau. In this letter we proposed a simple model, based
on  the role of cooperative effects, that is able to explain this result. In the whole paper we will consider a cross-linked elastomer of reticulation number $P$ (the mesh-size is $\lambda_{0} = aP^{\frac{1}{2}}$, where $a$ is the monomer size), in contact  with a flat neutral surface with N-mer grafted chains ($N>P$) of the same chemical constitution. The starting point of our study is the description of the penetration of a single grafted chain in a static rubber, made by O'Connor and McLeish \cite{OConnor93}. They assume that the chains can be confined in a slab between the elastomer and the flat surface without entangling with the elastomer (see fig.~\ref{f.1}). This slab should correspond to the first mesh of the elastomer and its width $\lambda$ be on the order of $\lambda_{0}$. They considered the case where the end of a grafted chain penetrates the elastomer at a distance $d$ from the grafting point. This leads to a partial penetration on only $m$ monomers, while the $n$ remaining monomers ($n = N - m$) are confined in the slab.

\begin{figure}
   \centering
   \includegraphics[clip=true, bb=2cm 12cm 27cm 20cm, width = 13cm]{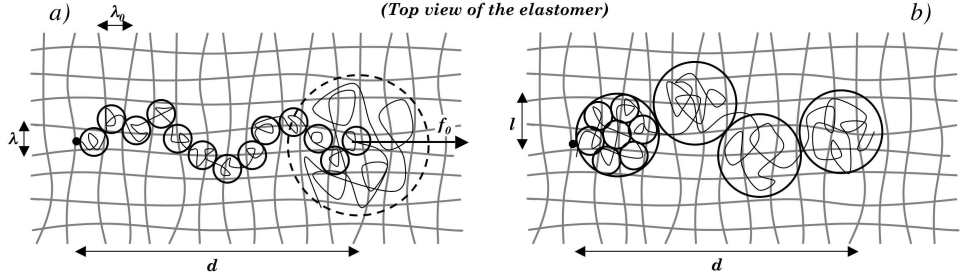}
\caption{a) Top view of a grafted chain that has partially entered the elastomer. b) The same chain confined in the slab when relaxing from the stretched conformation a).}
\label{f.2}
\end{figure}

As the swelling energy of the elastomer can easily be shown to be negligible for a single chain (and even for higher grafting densities \cite{Vilmin1}, see last section), the $m$ monomers 'feel' like in a melt of longer chains, and the free energy of the chain only contains the stretching and the confinement energies of the $n$ monomers \cite{Deutch94} :

\begin{equation}
\frac{F(n)}{kT} \simeq
\frac{3}{2} \frac{d^{2}}{a^{2}n}+\frac{3}{2} \frac{a^{2}n}{\lambda^{2}}
\label{eq1}
\end{equation}

when $d>\lambda$. Minimizing this expression with respect to $n$, we get the equilibrium value of $n$: $n_{eq} = \lambda d/a^{2}$ if $d$ is inferior to $d_{max} =a^{2}N/\lambda$, and the minimum free energy $F_{in}/kT = 3d/\lambda$. The $n_{eq}$ monomers constitute a stretched string of blobs of size $\lambda$ (see fig.~\ref{f.1} and ~\ref{f.2}a), which exerts a strong horizontal force $f_{0} = 3kT/ \lambda$ on the elastomer. If $d<\lambda$, the stretching force is not horizontal, and its projection on the horizontal axis is $f \simeq 3kT d/ \lambda^{2}$.
This partial penetration state is metastable as long as $d \geq 0$, but its life time is very long, for the chain has to go through a high energy state in order to reach another state of smaller $d$. In this hight energy state the whole chain is confined in the slab, it forms a string of flat blobs of size $l = a^{2}N/d$ if $d > aN^{\frac{1}{2}}$, and a blob of size $aN^{\frac{1}{2}}$ otherwise (see fig.~\ref{f.2}b), and the free energy is $F_{out} = F(N)$. Then, the time it takes for the chains to relax to another state is proportional to $\exp [(F_{out}-F_{in})/kT]$.

When the elastomer slides on the grafted surface, the chain can not relax to the state where $d=0$. The competition between the relaxation and the pull-out due to the elastomer sliding is thus key to the understanding of how a grafted chain can enhance friction.
In the next section we will calculate the relaxation time for a given $d$. Then, we will see how a sliding velocity $\mathbf{v}$ 
imposes a mean extension $d$ and a pull-out friction force
$\mathbf{f}$. In another sections we will consider interactions between several grafted chains, and show how it can modify the average friction force by grafted chain. Lastly, we will compare our results with recent experiments proceeded by Bureau and L\'eger \cite{Bureau04}.

\section{Relaxation time}

When the chain partially penetrates the elastomer and has $n$ monomers in the slab, the difference between its free energy and the minimum free energy is $(F(n) - F_{in})/kT = 3 a^{2}(n - n_{eq})^{2}/(2 n \lambda^{2})$. Then, setting $l = a^{2}n/\lambda$, we can consider the chain free end as a random walker diffusing in the potential $U(l) = 3 (l - d)^{2}/(2 l \lambda)$ ($0<l<d_{max}$) with a diffusion coefficient $D_{eff}(l)$. The relaxation time we are looking for is the mean first passage time at $l=d_{max}$ (that is when the chain retracts entirely in the slab, $n=N$), from the initial position $l=d_{max}-\lambda_{0}$ (that is the chain just penetrates the elastomer on a mesh size, $n=N-P$). This mean first passage time $\tau_{in}$ can be shown \cite{Gardiner}\cite{McL97} to be:

\begin{equation}
\tau_{in}(d) = \int_{d_{max}-\lambda_{0}}^{d_{max}} dx  \exp{[U(x)]} \int_{0}^{x} dx' \frac{e^{-U(x')}}{D_{eff}(x')}
\label{eq2}
\end{equation}

The inner integral is dominated by the region near $x'=d$ and can be approximated by $(D_{eff}(d)(2U"(d)/\pi)^{\frac{1}{2}})^{-1}$, where $U"(d) = 3/\lambda d$, and $D_{eff}(d) = 2a^{2}/\tau_{0}N$ if $\lambda=\lambda_{0}/2$, which we will assume here-after ($\tau_{0}$ is the monomer relaxation time). The outer integral can be approximated by $\lambda_{0} \exp[{U(d_{max})}]$, as $(U'(d_{max}))^{-1} > \lambda_{0}$. Then, the mean relaxation time at a given $d$ is:

\begin{equation}
\tau_{in}(d) \simeq 
\\ \left\{
\begin{array}{l}
\tau_{0}N\frac{\lambda^{2}}{a^{2}}\sqrt{\frac{\pi}{6}} \exp{\left[\frac{3}{2}\frac{(d_{max} -\lambda)^{2}}{a^{2}N}\right]} \quad , \quad \mathrm{if} \quad d < \lambda \\
\tau_{0}N\frac{\lambda^{2}}{a^{2}}\sqrt{\frac{\pi d}{6\lambda}} \exp{\left[\frac{3}{2}\frac{(d_{max} - d)^{2}}{a^{2}N}\right]} \quad , \quad \mathrm{if} \quad \lambda < d < d_{max}-\frac{1}{2}\sqrt{\frac{\pi}{6}}aN^{\frac{1}{2}} \\
\tau_{0}N\frac{\lambda}{a^{2}}\left(\frac{1}{2}\sqrt{\frac{\pi}{6}}aN^{\frac{1}{2}} + d_{max} - d \right)  , \mathrm{if} \quad d_{max}-\frac{1}{2}\sqrt{\frac{\pi}{6}}aN^{\frac{1}{2}} < d < d_{max}
\end{array}
\right.
\label{eq3}
\end{equation}

Notice that $\tau_{in}$ does not vanish at $d = d_{max}$, as the chain end can still explore the inside of the elastomer on the curvilinear distance $\sqrt{\pi/6}(aN^{\frac{1}{2}}/2)$. This calculation of the relaxation time notably differs from the one of Rubinstein \textit{et al.} \cite{Bubin93}\cite{Ajdari94}, and the result is smaller by a factor $(P/N)^{\frac{1}{2}}$ to $P/N$ when $d < d_{max}-\sqrt{\pi/6}(aN^{\frac{1}{2}}/2)$.
Another interesting time is the mean first passage time at $n=N-P$ from the initial position $n=N$, which is the time $\tau_{out}$ the chain spends entirely in the slab before to hop inside the elastomer. It corresponds to the diffusion time of the chain free end through the first mesh of the elastomer: $\tau_{out} = \tau_{0}P^{2}$. During this short time, the free end is driven back toward the grafting point as a result of the elastic shrink of the chain on the mean distance $6d \tau_{out}/ \tau_{0} N^{2}$, but it also diffuses horizontally on the distance $\lambda_{0}$ which is much bigger than $6d \tau_{out}/ \tau_{0} N^{2}$. Therefore, diffusion dominates.

\section{Sliding friction}

If the elastomer slides on the grafted surface at the velocity $\mathbf{v}$, a fully penetrating chain would be pulled-out of the elastomer and stretched in the sliding direction. But if $v \tau_{in} < d_{max}$, the chain will spontaneously relax and hop out of the elastomer before its complete stretching. Then, a permanent regime will settle, corresponding to cycles of hopping in and out, fixing an average value for $d$.
Every hopping out, the free end of the chain diffuses in the slab on the distance $\lambda_{0}$, and every hopping in this free end is shifted of  $v \tau_{in}(d)$ in the sliding direction. Then, the smallest possible value $d_{min}$ for $d$ is given by the relation $v \tau_{in}(d_{min}) = \lambda_{0}$. All the values between $d_{min}$ and $d_{max}$ can be explored by diffusion, but as $\tau_{in}(d)$ is a strongly decreasing function of $d$, the time averaged value of $d$ is approximately $d_{min} - \lambda_{0}(v \tau_{in}'(d_{min}))^{-1}$, with the standard deviation $\delta d = - \lambda_{0}(v \tau_{in}'(d_{min}))^{-1}$. The value of $\delta d \simeq a^{2}N/3(d_{max}-d_{min})$ varies from $\lambda_{0}/6$ when $d_{min} = \lambda_{0}$, to $aN^{\frac{1}{2}}/\sqrt{3}$ when $d_{min} =d_{max} - aN^{\frac{1}{2}}/\sqrt{3}$, and is thus always much smaller than $d_{min}$. Then, we can approximate the average value of $d$ by $d_{min}$, which gives :

\begin{equation}
v \tau_{in}(d) \simeq \lambda_{0}
\label{eq4}
\end{equation}

Equation \ref{eq4} has been already proposed by Rubinstein \textit{et al.} \cite{Bubin93}, but the prediction for $\tau_{in}(d)$ differs from ours (see eq.  \ref{eq3}). Now that we get $d$ as a function of $v$, we can express the force $f$ applied by a grafted chain on the elastomer for different velocities. Four velocity regimes can be brought out of this result :

If $v < v_{1} \simeq \lambda_{0}e^{-\frac{3}{2}\frac{a^{2}N}{\lambda^{2}}}/(\tau_{0} NP)$, then $d < \lambda$, the chain is almost fully relaxed, and $f \simeq f_{0}\frac{v}{v_{1}}$. For we consider cases where $a^{2}N \gg \lambda^{2}$, $v_{1}$ is extremely small.

If $v_{1} < v < v_{2} \simeq a/(\tau_{0} N^{\frac{3}{2}})$, then the chain is partially stretched, and as $\lambda < d < d_{max}$,  $f \simeq f_{0}$. In addition to this stretching force we shall take into account the Rouse friction of the elastomer on the whole chain \footnote{Even if the the part of the chain that is in the elastomer is relaxed, it is pulled-out at the velocity $v\lambda/\lambda_{0}$ in average.}, which is approximately $kT \tau_{0} N v / a^{2}$, much smaller than $f_{0}$ in this regime and the latter one. We note $v_{1}' \simeq v_{2}\sqrt{aN^{\frac{1}{2}}/\lambda}\exp[-3(d_{max}-aN^{\frac{1}{2}})^{2}/2a^{2}N]$ the velocity for which $d=aN^{\frac{1}{2}}$. Below $v_{1}' $ the chain relaxes forming a flat blob of size $aN^{\frac{1}{2}}$. The velocities $v_{1}$ and $v_{1}'$ are extremely small. When $v < v_{2}/2$, as $d < d_{max}-\frac{1}{2}\sqrt{\frac{\pi}{6}}aN^{\frac{1}{2}}$, $d$ is given by the relation

\begin{equation}
\frac{v}{v_{2}} \simeq \frac{1}{2}\sqrt{\frac{d_{max}}{d}} e^{-\frac{3}{2}\frac{a^{2}N}{\lambda^{2}}\left(1-\frac{d}{d_{max}}\right)^{2}}
\label{eq5}
\end{equation}

If $v_{2} < v < v_{3} = a/(\tau_{0} P^{\frac{3}{2}})$, then the chain is pulled-out faster than it can relax, $d = d_{max}$, and the friction force on the chain is $f_{0} + kT \tau_{0} N v / a^{2}$, where $kT \tau_{0} N v / a^{2}$ is no longer much smaller than $f_{0}$.

At last, if $v_{3} < v$, the Rouse friction dominates, and $d > d_{max}$.

These regimes have been depicted by Rubinstein \textit{et al.}, but the expressions of $v_{1}$ and $v_{2}$ differs, mainly because their estimation of $\tau_{in}$ differs from ours. Notice that the friction of the elastomer on the substrate is to be added to the friction of the chains, and that it could be partially screened out by the grafted layer \cite{Brown94}.

At this point it is important to notice that nothing prevents the chain orientation to fluctuate around the sliding direction\footnote{This was not possible in the situation Rubinstein \textit{et al.} studied, as the chain was dragged inside the elastomer.}. If the angle between the chain orientation and $\mathbf{v}$ is $\theta$, one should replace $v$ by $v cos\theta$ in equation \ref{eq4}, which really changes $d$ only for angles close to $\pi/2$ or $-\pi/2$. These fluctuations lower the effective pull-out friction force by grafted chains of a factor $2/\pi$.
Considering more than one grafted chain, one can foresee that a more important consequence of these fluctuations is that it allows the chains to entangle one with the other.

\section{Cooperative effects at higher grafting densities}

\begin{figure}
   \centering
   \includegraphics[clip=true, bb=2cm 12cm 27.5cm 19cm, width = 13cm]{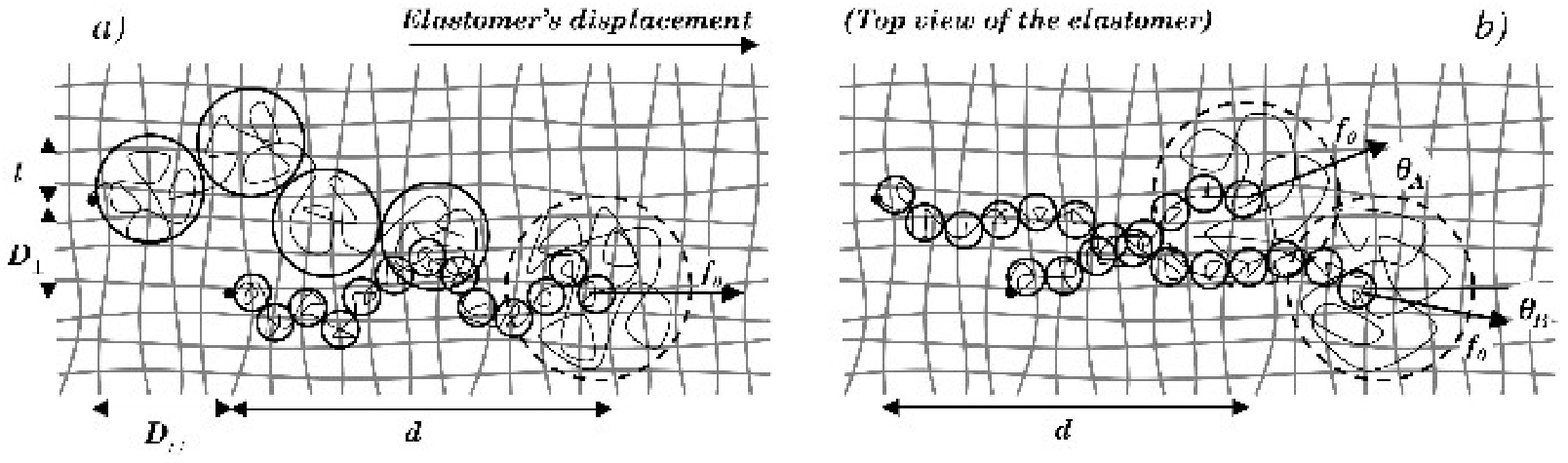}
\caption{Entanglement process between two grafted chains: a) The free end of the relaxed chain $B$ passes over and underneath the stretched part of the chain $A$. b) The chain $B$ recovers its plume conformation.}
\label{f.3}
\end{figure}

If two chains are grafted at a distance $D$ smaller than $d$, they can cross, and possibly entangle one with the other. Two situations can be distinguished. 
First, if the distance $D_{\bot}$ between the two grafting points perpendicularly to $\mathbf{v}$ (see fig.~\ref{f.3}a) is bigger than the size $l$ of the flat blobs, the entanglement can only form at the end of one of the chains, and will untie within approximately $(l/\lambda_{0})^{4}$ hopping in and out cycles (the time for the end of the chain to diffuse on the size $l$), whereas it took $(D_{\bot}/\lambda_{0})^{4}$ cycles to form. Thus, that kind of entanglement is fleeting and irrelevant.
If $D_{\bot}$ is smaller than $l$, then an entanglement can form relatively quickly and let the chains recover the same orientation as $\mathbf{v}$. Then, the stretching forces of the two chains equilibrate, and the entanglement slides toward the middle of the chains (see fig.~\ref{f.3}b). The effective friction force applied on the elastomer by those two chains is then $f_{A} + f_{B} = f_{0} (cos\theta_{A} + cos\theta_{B}) < 2f_{0}$. This entanglement can untie only if one of the chains free end reaches it. But as $d$ can not be smaller than $d_{min}$, this kind of entanglement has a quasi-infinite life time if the distance $D_{\Vert}$ between the two grafting points parallel to $\mathbf{v}$ is smaller than $d_{min}$. Therefore, we can assume that a chain entangle with all the chains that are grafted within the area $2l \times d_{min} \simeq 2ld = 2a^{2}N$ if $v_{1}' < v < v_{2}$. Then, if $\sigma$ is the grafting density scaled by $a^{2}$, we can roughly evaluate the average value of $cos\theta$ as

\begin{equation}
<cos\theta> \simeq \frac{\frac{d}{2\sigma N}}{\sqrt{\left(\frac{d}{2\sigma N}\right)^{2}+\left(\frac{l}{2}\right)^{2}}} = \frac{1}{\sqrt{1+\sigma^{2}\left(\frac{aN}{d}\right)^{4}}}
\label{eq6}
\end{equation}

This gives the pull-out friction by surface unit as a function of $\sigma$ :

\begin{equation}
\Sigma \simeq \frac{\sigma f_{0}}{a^{2}\sqrt{1+\sigma^{2}\left(\frac{aN}{d}\right)^{4}}} \simeq 
\\ \left\{
\begin{array}{l}
\, f_{0}\frac{\sigma}{a^{2}} \quad\quad , \quad \mathrm{if} \quad \sigma < \left(\frac{d}{aN}\right)^{2} \\
f_{0}\frac{d^{2}}{a^{4}N^{2}} \quad \;\; , \quad \mathrm{if} \quad \sigma > \left(\frac{d}{aN}\right)^{2} \\
\end{array}
\right.
\label{eq7}
\end{equation}

So, this rough model exhibit an interesting feature of elastomer-grafted surface friction: at low grafting densities ($\sigma < \sigma_{l} \simeq (d/aN)^{2}$) the friction force by surface unit increases linearly with $\sigma$, there is no interaction between grafted chains. At higher grafting densities ($\sigma > \sigma_{l}$) the friction force by surface unit saturates at the value $\Sigma_{l} = f_{0}d/aN$ (we don't consider here the friction of the elastomer on the substrate). The roughness of our evaluation of $<cos\theta>$ actually only allows us to give an evaluation of $\sigma_{l}$, give or take a multiplicative constant. Nevertheless, $\sigma_{l} \simeq (d/aN)^{2} = a^{2}/l^{2}$ is the grafting density beyond which the mean distance between grafted chains is smaller than $l$, and we can understand that entanglements produce an important orientation disorder beyond this limit.

The general tendencies for $\sigma_{l}$ correspond to the four velocity regimes we developed in the previous section.
If $v < v_{1}'$, then $l=aN^{\frac{1}{2}}$ and $\sigma_{l} = \sigma_{min} \simeq 1/N$, whereas if $v_{2}<v<v_{3}$, then $l=\lambda$ and $\sigma_{l} = \sigma_{max} \simeq 4/P$. When $v > v_{3}$, the Rouse friction dominates and should not be sensitive to entanglements. Using equation \ref{eq5} we can establish the relation between $v$ and $\sigma_{l}$ between $v_{1}' $ and $v_{2}/2$ :

\begin{equation}
\frac{v}{v_{2}} \simeq \frac{1}{2}\left(\frac{\sigma_{max}}{\sigma_{l}}\right)^{\frac{1}{4}} e^{-\frac{3}{2}\frac{a^{2}N}{\lambda^{2}}\left(1-\left(\frac{\sigma_{l}}{\sigma_{max}}\right)^{\frac{1}{2}}\right)^{2}}
\label{eq8}
\end{equation}

\begin{figure}
   \centering
     \includegraphics [clip=true, bb=2.5cm 7.5cm 27cm 14.5cm, width = 12cm]{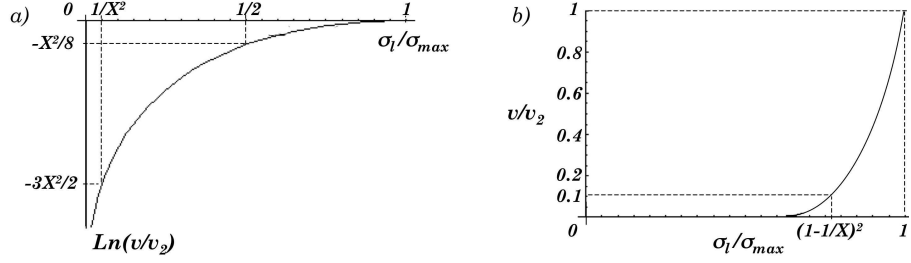}
\caption{Characteristic plots of the evolution of $\sigma_{l}/\sigma_{max}$ with $v/v_{2}$ ($X^{2} = a^{2}N/\lambda^{2}$).}
\label{f.4}
\end{figure}

The ratio $X^{2} = a^{2}N/\lambda^{2} \sim N/P$ is an important parameter characterizing the system; it gives the range of saturation grafting densities $\sigma_{max}/\sigma_{min} = X^{2}$. The parameter $X$ also drives the way $\sigma_{l}$ evolves within $(\sigma_{min}, \sigma_{max})$ while $v$ varies from $v_{1}'$ to $v_{2}$ (see fig.~\ref{f.4}). Indeed, the range of velocities for which $\sigma_{l}$ is bigger than $\sigma_{max}/2$ is given by $v(1/2)/v_{2} \simeq \exp\left[-X^{2}/8\right]$, which represents several decades when $X \gg 1$ (see fig.~\ref{f.4}a).  Another way to see the role played here by $X$ is to write the saturation grafting density corresponding to $v = v_{2}/10$: $\sigma_{l}(1/10)/\sigma_{max} \simeq (1-1/X)^{2}$ (all logarithmic factor being ignored, see fig.~\ref{f.4}b), which correspond to $d \simeq d_{max} - N^{\frac{1}{2}}$.

It is interesting to note that $\sigma_{max}$ and $\Sigma_{max}$ are independent of the chains length $N$; $\Sigma_{max}$ being the maximum friction enhancement one can obtain grafting polymer chains on the flat surface, and $\sigma_{max}$ the minimum grafting density one should use in order to reach $\Sigma_{max}$. 
The grafted chains length is though an important parameter as it fixes the velocity range over which $\sigma_{l} \sim \sigma_{max}$. One should use long chains ($a^{2}N \gg \lambda^{2}$) in order to have $\Sigma_{l}$ close to $\Sigma_{max}$ at very low velocities.
Nevertheless, the maximum grafting density one can experimentally reach is $1/N^{\frac{1}{2}}$, and one can show that beyond $\sigma^{*} \simeq P^{\frac{1}{10}}N^{-\frac{3}{5}}$ the grafted layer no longer interdigitates with the elastomer \cite{Vilmin1}; thus the maximum chain length one should use is $N \simeq P^{\frac{11}{6}}$.

\section{Results and experiment comparison}

Comparison with experiments is made harder by the fact that $\lambda$ is an unknown parameter assumed to be approximately $\lambda_{0}/2$. However, it can give insights into the validity of the model. Using a PDMS elastomer with reticulation number $P = 100$ and a grafted surface of $N = 1540$\cite{Bureau04} and $380$(unpublished studies) PDMS chains, Bureau \textit{et al.} systematically studied $\Sigma = f(\sigma)$ for sliding velocities from $0.3$ to $250\mu m.s^{-1}$, each time observing a friction force by grafted chain on the order of $kT/\lambda_{0}$ at low grafting densities, and a saturation of the friction at high grafting densities \cite{Bureau04}. 

For $N=1540$, they studied saturation at $v=0.3$, $10$, and $100\mu m.s^{-1}$, while $v_{2} \simeq 300\mu m.s^{-1}$. At $v=0.3\mu m.s^{-1}$, $\sigma_{l} \simeq 0.025 nm^{-2}$, at $v=10\mu m.s^{-1}$, $\sigma_{l} \simeq 0.035 nm^{-2}$, and at $100\mu m.s^{-1}$ the friction seems to saturate around $0.04 nm^{-2}$.
For $N=380$, they studied saturation at $v=10$, $100$, and $250\mu m.s^{-1}$, while $v_{2} \simeq 2000\mu m.s^{-1}$. At $v=10\mu m.s^{-1}$, $\sigma_{l} \simeq 0.015 nm^{-2}$, at $v=100\mu m.s^{-1}$, $\sigma_{l} \simeq 0.025 nm^{-2}$, and at $250\mu m.s^{-1}$, the friction seems to saturate at $\sigma_{l} \simeq 0.035 nm^{-2}$.

As $1/a^{2}P = 0.04nm^{-2}$, $\sigma_{max} \simeq 0.08nm^{-2}$ would fit with $\lambda \simeq 0.7\lambda_{0}$. Then, for $N=1540$, $X^{2} = a^{2}N/\lambda^{2} \simeq 30$. For the three velocities studied experimentally, expression \ref{eq8} gives $\sigma_{l}(0.3) \simeq 0.03 nm^{-2}$, $\sigma_{l}(10) \simeq 0.04 nm^{-2}$, and $\sigma_{l}(100) \simeq 0.055 nm^{-2}$.
For $N=380$, $X^{2} \simeq 8$, then, expression \ref{eq8} gives $\sigma_{l}(10) \simeq 0.01 nm^{-2}$, $\sigma_{l}(100) \simeq 0.02 nm^{-2}$, and $\sigma_{l}(250) \simeq 0.03 nm^{-2}$.
Thus, the model reasonably captures those experimental data, even if a slight misevaluation of $\lambda$ would induce consequent errors on $X^{2}$ and $\sigma_{l}(v)$. 

Although more data are needed to confirm the model, we think that it is the best candidate explaining the saturation of the friction : The first other possibility comes from the fact that increasing the grafting density can induce an increase of $\lambda$, and a decrease of $f_{0} \simeq kT/\lambda$. Indeed, we can estimate that $\lambda \simeq \lambda_{0}/2 +\sigma n_{eq}a$. Then, $\sigma f_{0} \simeq \sigma kT/(\lambda_{0}/2 +\sigma n_{eq}a)$, which would give a saturation grafting density equal to $P^{\frac{1}{2}}/n_{eq}$. This overestimates $\sigma_{l}$, and would give $\sigma_{l}$ as an decreasing function of $v$ and $N$, which is not the case experimentally.
The second possibility comes from the fact that when $\sigma > \sigma^{*} \simeq P^{\frac{1}{10}}N^{-\frac{3}{5}}$, the grafted chains no longer penetrate the elastomer because the swelling of the elastomer would not be negligible any more. But, again, this over estimate $\sigma_{l}$, and would give $\sigma_{l}$ as an decreasing function of $N$.

\section{Conclusion}

We have described here a model for the sliding friction of an elastomer on a flat grafted surface that allows us to understand the participation of grafted chains on friction at low and high grafting densities. This model describes the pull out process of the grafted chains and the formation of entanglements between grafted chains at high grafting densities. The general feature that follows from this is a linear relation between the pull out friction force by surface unit and the surface grafting density when $\sigma$ is low, and a saturation of the pull out friction force by surface unit beyond a limit grafting density $\sigma_{l}$, which is an increasing function of $v$ and $N$. The predictions for $\sigma_{l}$ are in good agreement with experimental results of Bureau \textit{et al.} which have been mainly conducted in the range $v < v_{2}$. For $v_{2} < v < v_{3}$, our model predicts that $\sigma_{l}$ simply varies like $1/P$; it would be very interesting to experimentally  check this prediction in the future.

We thanks Lionel Bureau and Liliane L\'eger for very interesting discussions and for letting to our knowledge unpublished results.

\end{document}